\documentstyle[12pt,epsfig]{article}

\begin{document}
\baselineskip=23pt


\begin{center}
{\Large \bf  Vacuum Energy Density and Cosmological Constant in dS
Brane-World}

\bigskip

Zhe Chang\footnote{changz@mail.ihep.ac.cn}, Shao-Xia
Chen\footnote{ruxanna@mail.ihep.ac.cn}
\\
{\em Institute of High Energy Physics,
Chinese Academy of Sciences} \\
{\em P.O.Box 918(4), 100039 Beijing, China}
\\
Xin-Bing Huang\footnote{huangxb@pku.edu.cn}
\\
{\em Department of Physics, Peking University,
 100871 Beijing, China}
\\
Hai-Bao Wen\footnote{wenhb@mail.ihep.ac.cn}\\
{\em Institute of High Energy Physics,
Chinese Academy of Sciences} \\
{\em P.O.Box 918(4), 100039 Beijing, China}
\end{center}

\bigskip
\bigskip

\centerline{\large Abstract} We discuss the vacuum energy density
and the cosmological constant of dS$_5$ brane world with a dilaton
field. It is shown that a stable AdS$_4$ brane can be constructed
and gravity localization can be realized. An explicit relation
between the dS bulk cosmological constant and the brane
cosmological constant is obtained. The discrete mass spectrum of
the massive scalar field in the AdS$_4$ brane is used to acquire
the relationship between the brane cosmological constant and the
vacuum energy density. The vacuum energy density in the brane
gotten by this method is in agreement with astronomical
observations.

\vspace{1.2cm}

\vspace{1.2cm}

\newpage

\section{Introduction}

In recent years, the old idea that placing our world on a domain
wall in a higher-dimensional bulk space has been used to explain
the hierarchy between the Planck scale and the electro-weak scale
in the four-dimensional effective field theory~[1-3].
After the notable papers of
Randall and Sundrum~\cite{RS1,RS3}, physicists have made progress
on using the idea of brane world to explain the cosmological
problems~\cite{CGKT,CGS}.

In the original Randall-Sundrum model, a flat $3$-brane
perpendicular to the fifth coordinate of the AdS$_5$ spacetime was
embedded. All matter and interactions except for gravity are
confined to the brane. Soon, the generalization of an AdS, flat or
dS brane in the AdS bulk~\cite{KSS1}, and a flat or dS brane in
the dS bulk were studied carefully\cite{I1}. The localization of
gravity in these models has been discussed
also~[4,9-12].
For the investigation of the brane world cosmology, the radion
field was introduced to stabilize the distance of
branes~\cite{CGRT,K1}. In some cases, the dilaton field was also
needed to get more predictions on
cosmology~[7,15-17].

The recent astronomical observations on Type Ia
supernovae~\cite{R1,P1} and the cosmological microwave
background~\cite{B1,Be} give more believable answers to several
long-existing problems, such as the cosmological constant,
flatness of space and existence of
inflation~[22-27].
However, the vacuum energy density calculated by quantum field
theory is much more larger than the possible observed
value~\cite{W1,CPT}. Various attempts have been made in trying to
solve the cosmological constant problem. Up to now, there still
hasn't a theory that can give a cosmological constant whose order
is the same as that of the observed value.  What is more puzzled,
from the theoretical point of view, is that why the observed
vacuum energy density is such a tiny value but
non-zero~[18-20,23-25].

In this paper, we discuss the vacuum energy density and the
cosmological constant of the dS$_5$ brane world with dilaton field
in a specially selected background metric. It is found that there
is not an AdS$_4$ brane solution in a dS$_5$ bulk without other
fields. Furthermore, one can not find an AdS$_4$ brane structure
by just introducing a dilaton field in the dS$_5$
bulk~\cite{KSS1,KSS2,CCH}. It is shown that both the dilaton field
and the special background metric are necessary to get a stable
AdS$_4$ brane in the dS$_5$ brane world. We study the localization
of gravity by expanding the traceless transverse component of the
perturbation of gravity in terms of the mass eigenstates of scalar
field in the brane. An explicit relation between the dS bulk
cosmological constant and the brane cosmological constant is
obtained. We solve exactly the massive scalar field equation in
the AdS$_4$ brane. The discrete mass spectrum of the massive
scalar field in the brane is used to acquire the relationship
between the brane cosmological constant and the vacuum energy
density. The vacuum energy density in the brane world gotten by
this method is in the same order with that of the observed value.

The paper is organized as follows. In section 2, we present the
general framework of the dS brane world with dilaton field.
Equations of motion for the gravity and dilaton field are derived.
Section 3 is devoted to finding a stable AdS$_4$ brane solution in
the dS brane world. The matching conditions on two sides of the
brane give a useful relation between two cosmological constants.
In section 4, localization of gravity is discussed by expanding
the traceless transverse component of the fluctuation of gravity
in terms of the mass eigenstates in the brane. We demonstrate that
the deviation of classical gravity from the Newton's law in the
brane is too small to be observed. In section 5, we calculate the
vacuum energy density of the brane by making use of exact
solutions of massive scalar fields. The obtained vacuum energy
density in the brane is in agreement with astronomical
observations. In section 6, we give conclusion and some remarks.

\section{Basic Setup and the Equations of Motion }

We consider a dS$_5$ spacetime with an AdS$_4$ brane embedding at
$y=0$. It is assumed first that there is only gravity in the bulk.
The action of the dS brane world can be written as
\begin{equation}
\label{action1} S=\int d^5
x\left[\sqrt{-G}\left(\frac{M^3}{2}R-\Lambda
\right)-\sqrt{-g}V\delta(y)\right]~,
\end{equation}
where $\Lambda$ ($>0$) and $M$ are the bulk cosmological constant
and five-dimensional fundamental scale, respectively. Following
Randall and Sundrum, we suppose the metric on the brane world is
as follows
\begin{equation}
\label{m}
 ds^2=e^{2A(y)}g_{\mu\nu}dx^\mu
dx^\nu+dy^2~,
\end{equation}
where $g_{\mu\nu}$ is the metric on the brane and takes the form
as\footnote{Here, for simplicity we define $(t\equiv
x^{0},\rho\equiv x^{1},\theta\equiv x^{2},\omega\equiv x^{3})$.
Five-dimensional suffices are denoted by capital Latin and the
four-dimensional one by the Greek ones.}
\begin{equation}
\label{g} g_{\mu\nu}={\rm diag}\left(-1,~
\frac{\cos^2{\sqrt{-\lambda}t}}{(1+\lambda{\rho}^2)^2},~
\frac{{\rho}^2\cos^2{\sqrt{-\lambda}t}}{1+\lambda{\rho}^2},~
\frac{{\rho}^2\sin^2{\theta}\cos^2{\sqrt{-\lambda}t}}{1+\lambda{\rho}^2}\right)~.
\end{equation}
Here $\sqrt{-\lambda}$ ($\lambda<0$) is the curvature of the AdS
brane.

The five dimensional Einstein equations for the above action read
\begin{equation}
\label{ein}
 \sqrt{-G}\left(R_{MN}-\frac{1}{2}G_{MN}R\right)= -
\frac{1}{M^3}\left[\Lambda \sqrt{-G}G_{MN}+\sqrt{-g}V
g_{\mu\nu}\delta^\mu_M\delta^\nu_N\delta(y)\right]~.
 \end{equation}
By making use of the ansatz (\ref{m}), we transform Eq.(\ref{ein})
into the form
\begin{equation}
\label{eq} \begin{array}{l} \displaystyle
6(A^\prime)^2+\frac{\Lambda}{M^3}=6\lambda e^{-2A}~,
\\[0.5 cm]
\displaystyle 3\lambda
e^{-2A}+3A^{\prime\prime}=-\frac{V}{M^3}\delta(y)~,
\end{array}
\end{equation}
where $^\prime$ denotes derivative with respect to $y$. It is
obvious that one can not find a nontrivial solution of the above
equations. Thus, we know there doesn't exist an AdS$_4$ brane
structure in this dS$_5$ spacetime with only gravity living in the
bulk.

To obtain a stable AdS$_4$ brane solution, we add a dilaton field
$\phi(y)$ in the dS$_5$ spacetime. The action of the system is of
the form
\begin{equation}
\label{s}
 S=\int
d^5x\left(\sqrt{-G}\left[R-\frac{4}{3}(\nabla\phi)^2-2\Lambda
e^{a\phi}\right] -2\sqrt{-g}\delta(y)Ve^{b \phi}\right)~.
\end{equation}
We suppose the spacetime metric as follows
\begin{equation}
\label{G}
 ds^2=e^{2A(y)}g_{\mu\nu}dx^\mu
dx^\nu+e^{2B(y)}dy^2~,
\end{equation}
where ${B(y)}$ is only a function of the fifth coordinate and
$g_{\mu\nu}$ takes the same form as Eq.(\ref{g}).

Equation of motion of the dilaton field reads
\begin{equation}
\label{V} \sqrt{-G}\left(\frac{8}{3}\nabla^2 \phi-2a\Lambda
e^{a\phi}\right)-2b\sqrt{-g}V\delta(y)e^{b\phi}=0~.
\end{equation}
The Einstein equation coupled with the dilaton field is of the
form
\begin{equation}
\label{E}
\begin{array}{rcl}
\displaystyle\sqrt{-G}\left(R_{MN}-\frac{1}{2}G_{MN}R\right)
&-&\displaystyle\frac{4}{3}\sqrt{-G}
\left[\nabla_M\phi\nabla_N\phi-\frac{1}{2}G_{MN}(\nabla\phi)^2\right]\\[0.5cm]
& &\displaystyle +\Lambda
e^{a\phi}\sqrt{-G}G_{MN}+\sqrt{-g}Ve^{b\phi}g_{\mu\nu}\delta^\mu_M\delta^\nu_N\delta(y)=0~.
\end{array}
\end{equation}
By making use of the ansatz for the metric (\ref{G}), we can
transform the equations of motion into the form
\begin{equation}
\label{d} \frac{32}{3}A^\prime\phi^\prime
e^{-2B}-\frac{8}{3}B^\prime\phi^\prime
e^{-2B}+\frac{8}{3}\phi^{\prime\prime}e^{-2B}-2a\Lambda
e^{a\phi}-2bV\delta(y)e^{b\phi}=0~,
\end{equation}
\begin{equation}
\label{44}
 -3\lambda e^{-2A+2B}+6(A^\prime)^2-3A^\prime
B^\prime+3A^{\prime\prime}+\frac{2}{3}(\phi^\prime)^2+\Lambda
e^{a\phi+2B}+V e^{b\phi}\delta(y)=0~,
\end{equation}
\begin{equation}
\label{55}
 -6\lambda
e^{-2A+2B}+6(A^\prime)^2-\frac{4}{3}(\phi^\prime)^2
e^{-2B}+\frac{2}{3}(\phi^\prime)^2+\Lambda e^{a\phi+2B}=0~.
\end{equation}
In the bulk, Eq.(\ref{d}) and Eq.(\ref{44}) reduce to
\begin{equation}
\label{db} \frac{32}{3}A^\prime\phi^\prime
e^{-2B}-\frac{8}{3}B^\prime\phi^\prime
e^{-2B}+\frac{8}{3}\phi^{\prime\prime}e^{-2B}-2a\Lambda
e^{a\phi}=0~,
\end{equation}
\begin{equation}
\label{44b}
 -3\lambda e^{-2A+2B}+6(A^\prime)^2-3A^\prime
B^\prime+3A^{\prime\prime}+\frac{2}{3}(\phi^\prime)^2+\Lambda
e^{a\phi+2B}=0~.
\end{equation}


\section{Solution and Relationship between Cosmological Constants}

To get an analytic solution of equations of motion (\ref{55}),
(\ref{db}) and (\ref{44b}), we assume that the following relations
are satisfied by fields in the system
\begin{equation}
\label{assume} A=\alpha\phi~,~~~B=\phi~,~~~a=-2~.
\end{equation}
Thus, Eq.(\ref{db}) becomes
\begin{equation}
\label{solve}
\frac{32}{3}\alpha(\phi^\prime)^2-\frac{8}{3}(\phi^\prime)^2+\frac{8}{3}
\phi^{\prime\prime}+4\Lambda=0~.
\end{equation}
It is easy to solve the above equation for the dilaton field
$\phi(y)$. In the case of $\alpha < \frac{1}{4}$, we acquire
$\phi^{\prime}(y)$ as follows
\begin{equation}
\label{phiprime}  \phi^{\prime}(y)=\left\{\begin{array}{l}
\displaystyle-D\tanh\left[(1-4\alpha)D(y+d_1)\right]~,~~~{\rm
for}~y>0\\[0.2cm]
\displaystyle-D\tanh\left[(1-4\alpha)D(y+d_2)\right]~,~~~{\rm
for}~y<0\end{array}\right.~,
\end{equation}
where $D=\sqrt{\frac{3\Lambda}{2(1-4\alpha)}}$. We present here a
solution of the form
\begin{equation}
\label{<} \phi=\left\{\begin{array}{l}
\displaystyle-\frac{1}{1-4\alpha}\ln\left\{
\cosh\left[(1-4\alpha)D(y+d_1)\right]\right\}+f_1~,~~~{\rm
for}~y>0\\[0.2cm]
\displaystyle-\frac{1}{1-4\alpha}\ln\left\{
\cosh\left[(1-4\alpha)D(y+d_2)\right]\right\}+f_2~,~~~{\rm
for}~y<0\end{array}\right.~,
\end{equation}
where $d_1,~d_2,~f_1$ and $f_2$ are constants which will be
determined by self-tuning.

Imposing the discontinuity of $\phi^\prime(y)$ at $y=0$ on
(\ref{d}) and (\ref{44}), we get the matching conditions as
follows
\begin{equation}
\label{b} \phi^\prime(0^+)-\phi^\prime(0^-)=\frac{3}{4}bV~,
\end{equation}
\begin{equation}
\label{a} \phi^\prime(0^+)-\phi^\prime(0^-)=-\frac{1}{3\alpha}V~.
\end{equation}
Thus, we have
\begin{equation}
\label{bv} b=-\frac{4}{9\alpha}~.
\end{equation}
The continuity of $(\phi^\prime)^2$ and discontinuity of
$\phi^\prime(y)$ at $y=0$ give that
\begin{equation}
\label{+-} \phi^\prime(0^+)=-\phi^\prime(0^-)~.
\end{equation}
From equations (\ref{+-}) and (\ref{<}), one can obtain
\begin{equation}
 d_1=-d_2\equiv d~.
\end{equation}
By making use of equation (\ref{a}), we can acquire $d$ as
\begin{equation}
\label{dv} d=\frac{1}{\sqrt{6\Lambda(1-4\alpha)}}\ln{\left|
\frac{1+F}{1-F} \right|}~,
\end{equation}
where $F=\frac{V}{3\alpha}\sqrt{\frac{1-4\alpha}{6\Lambda}}$~.

\begin{figure}[t]
\centerline{
\includegraphics[width=4in]{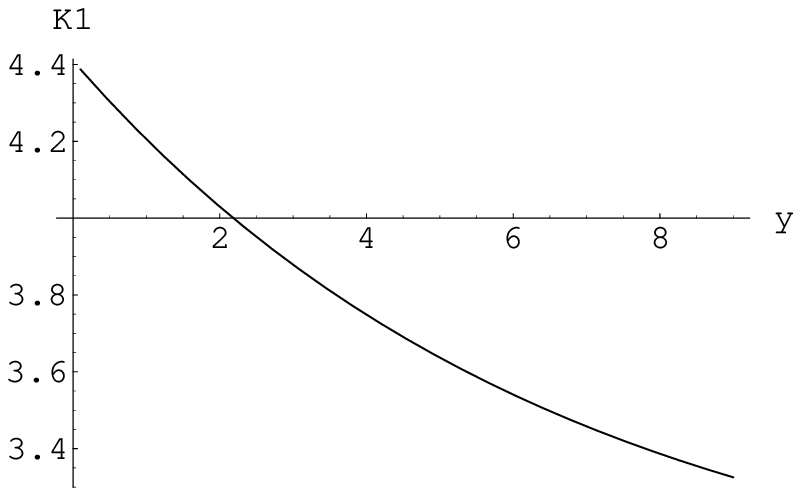}}
\caption{The behavior of the function~$K1$~(unit of the transverse
axis is~$10^{24}$ and of the vertical axis is~$10^{-51}$).}
  \label{k1}
\end{figure}

\begin{figure}[t]
\centerline{
\includegraphics[width=4in]{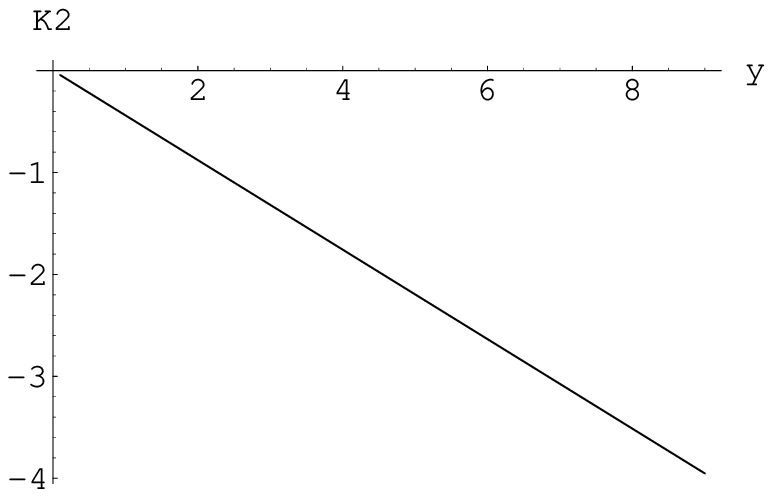}}
\caption{The behavior of the function~$K2$~(unit of the transverse
axis is~$10^{20}$ and of the vertical axis is~$10^{-55}$).}
  \label{k2}
\end{figure}

Furthermore, $f_1$ and $f_2$ can be determined by
$\phi(0^+)=\phi(0^-)=0$,
\begin{equation}
\label{fv}
f_1=f_2=-\frac{1}{2(1-4\alpha)}\ln\left(1-\frac{V^2}{9\alpha^2}\frac{(1-4\alpha)}{6\Lambda}\right)~.
\end{equation}
Inserting relations (\ref{assume}), (\ref{dv}) and (\ref{fv}) into
the equations of motion (\ref{55}) and (\ref{44b}), we get
\begin{equation} \label{44simple} -3\lambda
e^{(-2\alpha+2)\phi}+\left(\frac{2}{3}-6\alpha^2\right)(\phi^\prime)^2+\left(1-\frac{9}{2}\alpha\right)\Lambda=0~,
\end{equation}
\begin{equation}
\label{55simple} -6\lambda
e^{(-2\alpha+2)\phi}+\left(6\alpha^2-\frac{4}{3}e^{-2\phi}+\frac{2}{3}\right)(\phi^\prime)^2+\Lambda=0~.
\end{equation}
The values of $\phi^{\prime}(0)^{2}$ and $\phi(0)$ as well as the
equation (\ref{55simple}) give an explicit relation between the
bulk cosmological constant $\Lambda$ and the brane cosmological
constant $\lambda$,
\begin{equation}\label{54}
\Lambda=6\lambda-\frac{(6\alpha^2-\frac{2}{3})}{36\alpha^2}V^2~.
\end{equation}
At this stage, we can say that it is really possible to get an AdS
brane structure in dS spacetime by self-tuning the parameters. We
show in the following by a numerical method that (\ref{44simple})
and (\ref{55simple}) are satisfied when the cosmological constant
$\Lambda$, factor $\alpha$ and tension $V$ in the brane take some
specially selected values. For the purpose, one can introduce
functions $K1$~and~$K2$ as follows
\begin{equation}
\label{abc1} K1\equiv-3\lambda
e^{(-2\alpha+2)\phi}+\left(\frac{2}{3}-6\alpha^2\right)
(\phi^\prime)^2+\left(1- \frac{9}{2}\alpha\right)\Lambda~,
\end{equation}
\begin{equation}
\label{abc2} K2\equiv-6\lambda
e^{(-2\alpha+2)\phi}+\left(6\alpha^2-
\frac{4}{3}e^{-2\phi}+\frac{2}{3}\right) (\phi^\prime)^2+\Lambda~.
\end{equation}

Recent observations of Type Ia supernovae and the cosmic microwave
background indicate that our universe is dominated by a positive
cosmological constant~[18-21,25,26].
So
it is reasonable to assume that the current age of Universe times
the light speed equals the radius of the dS$_5$ spacetime.
Therefore, we have

\begin{equation}
\label{radius} \frac{1}{\sqrt{\Lambda}}=\frac{c}{H_0}
\end{equation}
where $H_0$ is present-day value of the Hubble expansion rate.
Namely, $\Lambda \simeq 5.89\times 10^{-53}{\rm m}^{-2}$. We make
the computer program to draw different plots of $K1$ and $K2$ by
automatically selecting different values of the parameters
$\alpha$, $V$ and $d$. Our program give the following best fit
values of $\alpha$, $V$ and $d$
\begin{equation}
\label{paraval}
\alpha=0.249,~~V=2.99\times 10^{-25},~~d=4.94\times
10^{27}{\rm m}.
\end{equation}
In figure 1 and figure 2 ,we show plots of $K1$ and $K2$ as
functions of $y$. It is clear that $K1$ and $K2$ are almost zero
in the whole area of $\displaystyle|y|<\frac{1}{\sqrt{\Lambda}}$.
We note that $K1$ and $K2$ deviate from zero very fast when
$|y|\rightarrow{d}$. Possible reason of the fact is that the
radius of five dimensional dS spacetime, the cosmological constant
and parameter $d$ satisfy the relation
\begin{equation}
\label{relation1}
\displaystyle \frac{1}{\sqrt{\Lambda}}~ \sim
\frac{1}{\sqrt{-\lambda}}~ \sim \frac{d}{10}~.
\end{equation}
So $|y|\rightarrow{d}$ is out of the dS spacetime. We can draw a
conclusion that the equations (\ref{44simple}) and
(\ref{55simple}) can be satisfied very well within the interval
$\displaystyle|y|<\frac{1}{\sqrt{\Lambda}}$ with the fine-tuned
parameters.

\section{Localization of Gravity}

To demonstrate the localization of gravity of AdS$_4$ brane in
dS$_5$ bulk with dilaton field, we rewrite the spacetime metric as
follows
\begin{equation}
\label{lg0a}
 ds^2=e^{2A(y)}(-dt^{2}+\gamma_{i}(t)^{2}\delta_{ij}dx^{i}
dx^{j})+e^{2B(y)} dy^{2}~,~~~~i,j=1,2,3,
\end{equation}
where
\begin{equation}
\label{lg0b}
\gamma_{1}=
\frac{\cos{\sqrt{-\lambda}t}}{(1+\lambda{\rho}^2)}~,~~~~\gamma_{2}
=\frac{{\rho}\cos{\sqrt{-\lambda}t}}{\sqrt{1+\lambda{\rho}^2}}~,~~~~
\gamma_{3}
=\frac{{\rho}\sin{\theta}\cos{\sqrt{-\lambda}t}}{\sqrt{1+\lambda{\rho}^2}}
~.
\end{equation}

Consider the perturbed metric $h_{ij}$ in the form\cite{bgoy02}
\begin{equation}
\label{lg0c}
 ds^2=e^{2A(y)}\left[-dt^{2}+\gamma_{i}(t)^{2}(\delta_{ij}+h_{ij}(x^{i}))dx^{i}
dx^{j}\right]+e^{2B(y)} dy^{2}~.
\end{equation}
We are interested in the localization of the traceless transverse
component of $h_{ij}$, which corresponds to the graviton of the
perturbation in the brane. It satisfies $D_ih^{ij}=0$ and
$h_i^i=0$. Here $D_i$ denotes the covariant differential with
respect to the space metric ${\gamma_{i}^{2}(t)}\delta_{ij}$. The
traceless transverse component of $h_{ij}$ is denoted by $h$ for
simplicity. $h$ satisfies the following
equation\cite{bgoy02,lan00}
\begin{equation}
\label{lg1}
\frac{1}{\sqrt{-G}}\frac{\partial}{\partial x^{M}} \left(
\sqrt{-G}G^{MN}\frac{\partial}{\partial x^{N}}h\right)=0~.
\end{equation}
It should be noticed that this is equivalent to a five dimensional
free scalar field equation. In our framework, the metric is
diagonal and Eq.(\ref{lg1}) can be reduced by expanding $h$ in
terms of four dimensional continuous mass eigen states
\begin{equation}
\label{lg2} h=\int dm ~\Phi_{m}(t,\vec{x}) ~\Psi(m,y)~,
\end{equation}
where $\Phi_{m}(t,\vec{x})$ is the mass eigen states of the
four-dimensional scalar field\footnote{Here $\vec{x}$ denotes
$(x^{1},x^{2},x^{3})=(\rho,\theta,\omega)$.}
\begin{equation}
\label{lg3} \frac{1}{\sqrt{-g}}\frac{\partial}{\partial x^{\mu}}
\left( \sqrt{-g}g^{\mu\nu}\frac{\partial}{\partial
x^{\nu}}\Phi_{m}(t,\vec{x})\right)=m^{2}\Phi_{m}(t,\vec{x})~.
\end{equation}
It is not difficult to obtain the equation for $\Psi(m,y)$ from
Eq.(\ref{lg1}) and Eq.(\ref{lg3})
\begin{equation}
\label{lg4}
\Psi^{\prime\prime}+(4A^{\prime}-B^{\prime})\Psi^{\prime}+m^{2}
{e^{2B-2A}}\Psi=0~.
\end{equation}

\begin{figure}[t]
\centerline{
\includegraphics[width=4in]{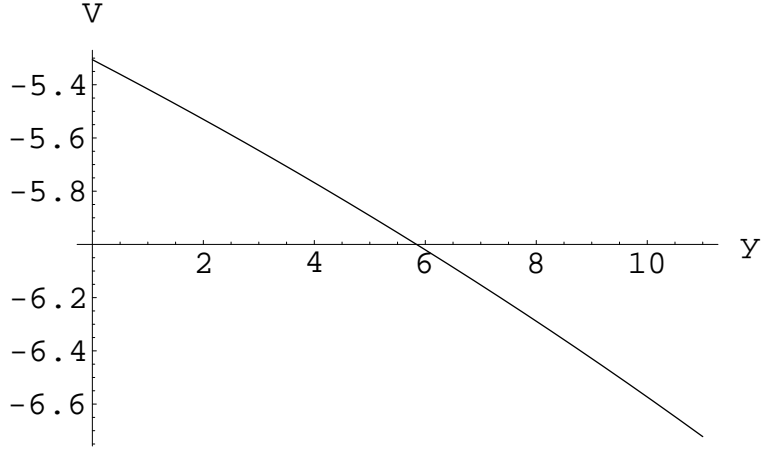}}
\caption{Potential tension $V$~(unit of the transverse axis
is~$10^{23}$ and of the vertical axis is~$10^{-5}$).}
  \label{pot}
\end{figure}
\begin{figure}[t]
\centerline{
\includegraphics[width=4in]{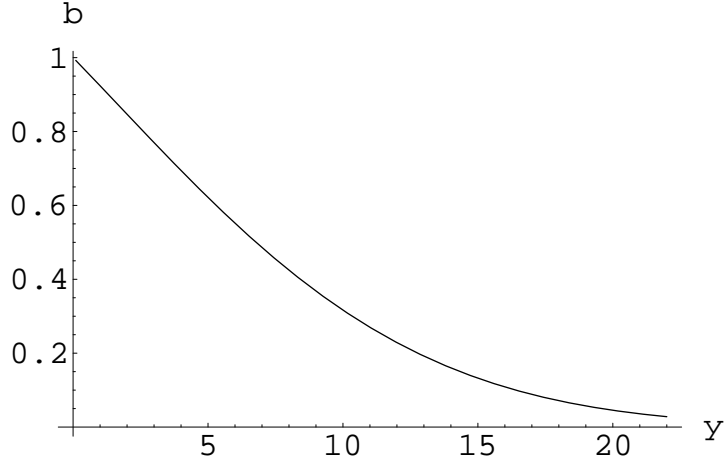}}
\caption{The behavior of the coordinate transformation
$\frac{\partial z}{\partial y}$~(unit of the transverse axis
is~$10^{24}$ and of the vertical axis is~$10^{1}$).}
  \label{local}
\end{figure}

Before considering the solution of Eq.(\ref{lg4}), we find that
the equation can be rewritten in a "supersymmetric" form by
introducing transformations $\Psi=e^{-\frac{3}{2}\alpha B}u(z)$
and $b\equiv \displaystyle \frac{\partial z}{\partial
y}=e^{(1-\alpha)B}$. Thus, Eq.(\ref{lg4}) becomes
\begin{equation}
\label{lg4+}
Q^{\dag}Qu(z)=(-\partial_{z}-\frac{3}{2}\alpha\frac{\partial
B}{\partial z})(\partial_{z}-\frac{3}{2}\alpha\frac{\partial
B}{\partial z})u(z) =m^{2}u(z)~.
\end{equation}
This equation demonstrates that the eigenvalue of mass should be
non-negative, i.e. no tachyon in the brane. Then the zero mode
$m=0$ is the lowest state which would be localized on the brane.
Furthermore, we can rewrite the equation for $u(z)$ as the
following one-dimensional Schr\"{o}dinger-like equation in the
$y$-direction with the eigenvalue $m^{2}$
\begin{equation}
\label{lg5} [-\partial^{2}_{z}+V(z)]u(z) =m^{2}u(z)~,
\end{equation}
where the potential $V(z)$ is determined by $B(z)$ as
\begin{equation}
\label{lg6} V(z)=\frac{3}{2}\alpha \left[\frac{3}{2}\alpha
{\left(\frac{\partial B}{\partial z}
\right)}^{2}+\frac{\partial^{2} B}{\partial z^{2}}\right]~.
\end{equation}
The localization is seen by solving Eq.(\ref{lg5}). The
localization of gravity can be concluded by the following two
conditions: First, the potential $V(z)$ is volcano-like and
contains a $\delta$-function attractive force at the brane
position to trap the zero-mode of the bulk graviton; Second, There
exist a normalizable state for the wave function $u_{0}(z)$ of
$m=0$ eigenvalue.

Figure 4 indicates that $b(y)$ could be treated as a linear
function in physicist's interested area. By the approximation, the
potential $V(z)$ is of the form
\begin{equation}
\label{lg60}
\begin{array}{rcl}
 V(z)&=&\displaystyle \frac{3}{2}\alpha
 \left(\frac{5}{2}\alpha-1\right)
D^{2}e^{-2\gamma B}\tanh^{2}\left[(1-4\alpha)D(z+{\rm sgn}(z)d)\right]\\
& & -\displaystyle \frac{3}{2} \alpha e^{-2\gamma B}\left \{
\frac{3\Lambda}{2} {\rm sech}^{2} \left[(1-4\alpha)D(z+{\rm
sgn}(z)d)\right]\right\} -\displaystyle\frac{V}{4}e^{-2\gamma
B}\delta(z)~.
\end{array}
\end{equation}
The $\delta$ function term in $V(z)$ guarantees that there exists
an independent unitary bound state solution
\begin{equation}
\label{lg7} u_{0}(z)=\frac{1}{\sqrt{L}}~e^{-|z|/L}~,
\end{equation}
where $L~(=\displaystyle \frac{4}{V})$ is the normalization
constant. The figures 3 and 4 show that the potential (\ref{lg60})
is a Volcano type. The perturbation of gravity with Volcano
potential has been studied extensively\cite{bgoy02,kr01a,kr01b}.
Therefore, we can say that the gravity of the dS$_5$ brane world
with dilaton field in a specially selected background metric is
well localized. That is to say, the deviation of the classical
gravitation from the Newton's law in the brane can be omitted.

\section{Vacuum Energy Density and Cosmological Constants}

The vacuum energy density in the AdS$_4$ brane can be calculated
by summing up the zero point energy of  different massive scalar
fields living in the brane. The equation of motion for a massive
scalar field in AdS$_4$ brane is of the form~\cite{CH,CGG}

\begin{equation}
\label{KG}
\begin{array}{l}
\displaystyle\left[\frac{1}{\cos^3{\sqrt{-\lambda}t}}\frac{\partial}{\partial
t} \left(\cos^3{\sqrt{-\lambda}t}\frac{\partial} {\partial t
}\right)-\frac{(1+\lambda\rho^2)^2}{\rho^2\cos^2{\sqrt{-\lambda}t}}
\frac{\partial}{\partial\rho}\left(\rho^2\frac{\partial}{\partial
\rho}\right)\right.\\[1cm]
\displaystyle~~~~~~~~~~~\left.-\frac{1+\lambda\rho^2}{\rho^2\cos^2{\sqrt{-\lambda}t}}
\left(\frac{1}{\sin\theta} \frac{\partial}
{\partial\theta}\left(\sin\theta\frac{\partial}{\partial\theta}\right)-
\frac{1}{\sin^2\theta}\frac{\partial^2}{\partial\phi^2}\right)
+m_{0}^2\right]\Phi(t ;\rho,\theta,\phi)=0~.
\end{array}
\end{equation}
The equation of motion can be solved exactly by the method of
variables separation. Solutions can be written as follows
\begin{equation}
\label{solution}
\begin{array}{l}
\Phi_{NIlm}(t;\rho,\theta,\phi) \propto U_{Nl}(\rho){\left( \cos
\sqrt{-\lambda} t \right)}^{-1} P_I^N\left(\sin {\sqrt{-\lambda}t
}\right)Y_{lm}(\theta,\phi)~,
\end{array}
\end{equation}
where $P_I^N(\sin\sqrt{-\lambda}t)$, with
$I={-\frac{1}{2}+\sqrt{\frac{1}{4}-\frac{m_{0}^2}{\lambda}+2}}~,
N=\sqrt{1+\frac{k^2}{\lambda}}~$, are associated Legendre
functions, and $U(\rho)$, the radial part of the wave function,
\begin{eqnarray}
\label{radialfunction}
\nonumber
U(\rho)=&C&(\sqrt{-\lambda}\rho)^{l}\left(1+\lambda{\rho^2}\right)^
{\frac{1}{2}+\frac{1}{2} \sqrt{1+\frac{k^2}{\lambda}}} \\
 &\times&_{2}{F_{1}\left(\frac{1}{2}(l+\sqrt{1+\frac{k^2}{\lambda}}+2),
\frac{1}{2}(l+\sqrt{1+\frac{k^2}{\lambda}}+1),l+\frac{3}{2};-\lambda{\rho^2}\right)}~,
\end{eqnarray}
here $C$ is the normalization constant.

The natural boundary condition for $P_I^N(\sin(\sqrt{-\lambda}t)
)$ on $\sin(\sqrt{-\lambda}t)= \pm 1$
 requires that $I,~ N$ to be integers.
This gives the discrete mass spectrum of the scalar fields in
AdS$_4$ brane~\cite{CGG},
\begin{equation}\label{mass}
\begin{array}{l}
\displaystyle -\frac{m_{0}^2}{\lambda}+2=I(I+1)~,\\[0.4cm]
\displaystyle 1+\frac{k^2}{\lambda}=N^2~,\quad  |N|\leq{I}~.
\end{array}
\end{equation}

The vacuum energy density on AdS$_4$ brane is obtained~\cite{CH}
\begin{eqnarray}
\label{cosm}
\langle\rho\rangle=\frac{1}{8{\pi}^2}\mid \lambda \mid
\frac{(E_{\rm Planck})^2}{\hbar c}~,
\end{eqnarray}
where $E_{\rm Planck}$ is the Planck energy.

The relation between the vacuum energy density and the
cosmological constant in the 5-dimensional spacetime can be
expressed as
\begin{equation}
\label{relation2}
\langle\rho\rangle=\frac{1}{8{\pi}^2}
\frac{(E_{\rm Planck})^2}{\hbar c}\bigg[
(\frac{1}{324\alpha^2}-\frac{1}{36})V^2-\frac{\Lambda}{6}\bigg]~.
\end{equation}

The vacuum energy density takes value
\begin{eqnarray}
\label{density}
\langle\rho\rangle=7.66\times 10^{-10}{\rm erg\cdot cm^{-3}}~.
\end{eqnarray}
The vacuum energy density gotten here is consistent with the
astronomical observations~\cite{B1,ZD,W1,C1,GZ}.

\section{Conclusion and Remarks}

The tiny vacuum energy density has puzzled physicists for nearly a
century. The recent astronomical observations on
supernovae~\cite{R1,P1} and CMB~\cite{B1,Be} show that about two
third of the world energy is contributed by a small positive
cosmological constant. The most simple cosmology model is an
asymptotic dS spacetime which has been discussed widely in the
literature~[37-39].
In this paper, we have
interpreted the tiny positive cosmological constant as the
curvature of a dS$_5$ brane world.

It is well known that the calculated vacuum energy density by
quantum field theory is much more larger than the possible
observed value~\cite{W1,CPT}. Various attempts have been made in
trying to solve the cosmological constant problem. To the end of
getting a comparable value of vacuum energy density with the
astronomical observations, we have tried to construct a AdS$_4$
brane in the dS$_5$ spacetime. It was shown that a dilaton field
is needed to get a stable AdS brane solution in the dS$_5$
spacetime. An explicit relation between the dS bulk cosmological
constant and the AdS brane cosmological constant has been
obtained. The discrete mass spectrum in AdS$_4$ brane was used to
acquire vacuum energy density. The cosmological constant in the
brane world gotten by this way is in the same order with the
astronomical observations.

{\bf Acknowledgement:} One of us (Z. Chang) would like to thank M.
Li, Y. S. Wu and C. J. Zhu for useful discussion. The work is
supported partly by the Natural Science Foundation of China.

\end{document}